\documentstyle[prl,aps,epsfig,floats]{revtex}

\newcommand\beq{\begin{equation}}
\newcommand\eeq{\end{equation}}
\newcommand\bea{\begin{eqnarray}}
\newcommand\eea{\end{eqnarray}}

\begin{document}

\draft

\textheight=23.8cm
\twocolumn[\hsize\textwidth\columnwidth\hsize\csname@twocolumnfalse\endcsname

\title{\bf Comment on ``Low-dimensional Bose liquids: beyond the \\
Gross-Pitaevskii approximation''}
\author{\bf R. K. Bhaduri$^1$ and Diptiman Sen$^2$}
\address{\it ${}^1$Department of Physics and Astronomy \\
McMaster University \\
Hamilton, Ontario, Canada L8S 4M1 \\
$^2$ Centre for Theoretical Studies \\
Indian Institute of Science \\
Bangalore 560012, India} 

\maketitle

\begin{abstract}
This is a comment on the work of Kolomeisky et al., Phys. Rev. Lett. 85, 
1146 (2000). We point out that they are using the wrong form of the energy 
functional for one-dimensional fermions. We point out two possible forms of 
the energy functional, both of which can be derived from first principles but 
using different methods. One is obtained from the collective field theory 
method, while the other is derived from the extended Thomas-Fermi method. 
These two forms of the energy functional do not support the soliton solutions 
which are obtained by Kolomeisky et al.
\end{abstract}

\vskip .5 true cm


\vskip.5pc
]
\vskip .5 true cm

The Gross-Pitaevskii (GP) mean-field theory replaces the bosonic field 
operator by a classical field $\Phi ({\bf r},t)$ \cite{pita}. 
This approach has been highly successful in describing a dilute gas of 
trapped bosonic atoms \cite{string}. However, some authors 
have pointed out that the usual potential energy $|\Phi|^4$ which arises from 
the zero-range pseudopotential in three dimensions, needs to be modified in 
lower dimensions. Kolomeisky {\it et al.} \cite{kolo} have 
proposed replacing the quartic term by a $|\Phi |^6$ term in one dimension. 
This can be motivated by considering a one-dimensional Bose gas in which 
particles interact pair-wise via a repulsive $\delta$-function potential. This 
is the Lieb-Liniger model which is exactly solvable \cite{lieb}. The model 
has only one dimensionless parameter, namely, $g =
\hbar^2 \rho / m u_0$, where $\rho =|\Phi|^2$ is the density, $u_0$ is the 
strength of the $\delta$-function interaction, and $m$ is the particle mass. 
In the limit of $g \rightarrow 0$ (called the dilute approximation), this 
system is equivalent to a gas of noninteracting fermions whose energy density 
is given by $\pi^2 \hbar^2 \rho^3 /6m$. Kolomeisky {\it et al.} use this 
energy density to study a dilute Bose gas and to obtain a 
stationary soliton solution. 

Our main criticism of their work is that they are using the wrong form of the 
{\it second derivative} terms in the energy functional for one-dimensional 
fermions, as indicated below. 
Since a dilute Bose gas with repulsive interactions in one dimension is 
equivalent to a system of noninteracting fermions, one way to proceed is 
to use the collective field theory (CFT) method to derive the Hamiltonian 
in terms of the density variables. This was found long ago \cite{cft}. In 
the absence of an external potential, the energy density is given by
\beq
H ~=~ \frac{\hbar^2}{2m} ~[~ \rho (\frac{\partial \theta}{\partial x})^2 ~
+~ \frac{\pi^2}{3} ~\rho^3 ~] ~,
\label{ham}
\eeq
where the density and phase fields $\rho$ and $\theta$ are related to the
order parameter field by $\Phi = {\sqrt \rho} e^{i\theta}$. 
The equations of motion for this system are easily obtained \cite{sen},  
and these do not support any solutions in which $\rho$ is 
time-independent and {\it inhomogeneous}. This seems to contradict
the fact that a system of noninteracting fermions can have a time-independent 
and inhomogeneous density profile \cite{gira}, for instance, 
in the vicinity of a hard wall. 
The exact density profile in both those cases has damped oscillations whose 
wavelength is of the order of the interparticle separation. This shows a 
limitation of the CFT method; the CFT Hamiltonian given in Eq. (\ref{ham}) 
gives the correct description of a system of noninteracting fermions only for 
long-wavelength density fluctuations \cite{sen}. A different derivation of the 
kinetic energy functional for noninteracting fermions placed in an external 
potential is based on the extended Thomas-Fermi method \cite{brack}. In one 
dimension, to order $\hbar^2$, this gives the inhomogeneous terms in the 
kinetic energy density to be $\hbar^2/2m (- \rho^{\prime 2} /12
\rho + \rho''/3 )$. This is different, in both sign and magnitude, from the 
term used by Kolomeisky {\it et al.}, and has no solitonic solutions. In 
contrast to this, the Hamiltonian of Kolomeisky {\it et al.} does support 
stationary solutions with inhomogeneous densities such as solitons. However, 
this does not by itself justify the addition of the term $\hbar^2 \rho^{\prime 
2} /8m \rho$ to the Hamiltonian (\ref{ham}) which describes noninteracting
{\it fermions}.

Our second comment on the work of Kolomeisky {\it et al.} is that the dilute 
approximation is not obtained in any of the Bose condensates studied so far 
\cite{string}. In the experimental systems, the equivalent of the parameter 
$g$ is $N a /a_{HO}$, where $a$ is the scattering length, and $a_{HO}$ is the 
harmonic confinement length. This equivalence follows because the particle 
density $\rho$ in the center of the trap is of the order of $N/a_{HO}$, 
while the scattering length $a$ is proportional to the strength of the 
pseudopotential in three dimensions and is therefore analogous to $u_0$ in 
the one-dimensional problem. Ref. \cite{string} states that $N a/a_{HO}$ 
typically goes from a number of order $1$ to several thousands, because 
$a/a_{HO}$ is usually of the order of $10^{-3}$ while $N$ typically goes from 
$10^3$ to $10^6$. Hence the experimental systems cannot be considered to be 
dilute in the Lieb-Liniger sense, and the mapping from interacting bosons to 
noninteracting fermions in one dimension is not valid for such systems. 

This research was partially supported by NSERC of Canada.
\vskip .6 true cm

\end{document}